\documentstyle[prl,aps,multicol]{revtex}
\begin{document}

\draft

\title{Skyrmions in Spinor Bose-Einstein Condensates}

\author{H.T.C. Stoof}
\address{Institute for Theoretical Physics,
         University of Utrecht, Princetonplein 5, \\
         3584 CC Utrecht,
         The Netherlands} 

\maketitle

\begin{abstract}
We show that spinor Bose-Einstein condensates not only have line-like vortex 
excitations, but in general also allow for point-like topological 
excitations, i.e., skyrmions. We discuss the static and dynamic properties of 
these skyrmions for spin-1/2 and ferromagnetic spin-1 Bose gases.
\end{abstract}


\begin{multicols}{2}
{\it Introduction.} --- An understanding of quantum magnetism is important for a 
large number of phenomena in physics. Three well-know examples are 
high-temperature superconductivity, quantum phase transitions and the quantum 
Hall effect. Moreover, it appears that magnetic properties will also be very 
important in another area, namely Bose-Einstein condensation in trapped atomic 
gases. The latter has come about because of two independent experimental 
developments. The first development is the realization of an optical trap for 
atoms, whose operation no longer requires the gas to be doubly spin-polarized 
\cite{MIT1,MIT2}. The second development is the creation of a two-component Bose 
condensate \cite{JILA1}, which by means of rf-fields can be manipulated so as to 
make the two components essentially equivalent \cite{JILA2}. As a result the 
behavior of both spin-1 and spin-1/2 Bose gases can now be experimentally 
explored in detail.

Theoretically, the ground-state structure of these gases has recently been 
worked 
out by a number of authors \cite{jason1,OM,nick,jason2} and also the first 
studies of the line-like vortex excitations have appeared \cite{jason1,Yip}. 
However, an immediate question that comes to mind is whether the spin degrees of 
freedom allow for other topological excitations that do not have an analogy in 
the case of a single component Bose condensate. It is the main purpose of this 
Letter to show that the answer to this question is in general affirmative. In 
particular, we show that spinor Bose-Einstein condensates have so-called 
skyrmion excitations, which are topological nontrivial point-like spin textures 
\cite{Skyrme}. Having done so, we then turn to the investigation of the 
precise texture and the dynamics of these skyrmions. 

{\it Topological considerations.} --- To find all possible topological 
excitations 
of a spinor condensate, we need to know the full symmetry of the macroscopic 
wavefunction  
$\Psi({\bf x}) \equiv \sqrt{n({\bf x})} \zeta({\bf x})$, where 
$n({\bf x})$ is the total density of the gas, $\zeta({\bf x})$ is a normalized 
spinor that determines the average local spin by means of 
$\langle {\bf S} \rangle({\bf x}) 
    = \zeta^*_{a}({\bf x}) {\bf S}^{ab} \zeta_{b}({\bf x})$, 
and ${\bf S}$ are the usual spin matrices obeying the commutation relations
$[S_{\alpha},S_{\beta}] 
                    = i {\epsilon_{\alpha\beta}}^{\gamma} S_{\gamma}$.
Note that here, and in the following, summation over repeated indices is always 
implicitly implied. From the work of Ho \cite{jason1} we know that in the case 
of spin-1 bosons we have to consider two possibilities, since the effective 
interaction between two spins can be either antiferromagnetic or ferromagnetic. 
In the antiferromagnetic case the ground-state energy is minimized for 
$\langle {\bf S} \rangle({\bf x}) = {\bf 0}$, which implies that the parameter 
space for the 
spinor $\zeta({\bf x})$ is only $U(1) \times S^2$ because we are free to 
chooce both its overall phase and the spin quantization axis. In the 
ferromagnetic case the energy is minimized for 
$|\langle {\bf S} \rangle({\bf x})| = 1$ 
and the parameter space corresponds to the full rotation group $SO(3)$. Using
the same arguments, we find that for spin-1/2 bosons the order parameter space
of the ground state is always equivalent to $SU(2)$ \cite{scat}.

What do these results tell us about the possible topological excitations? For
line-like defects or vortices, we can assume $\zeta({\bf x})$ to be
independent of one direction and the spinor represents a mapping from a
two-dimensional plane into the order parameter space. If the vortex is singular 
this
will be visible on the boundary of the two-dimensional plane and we need to
investigate the properties of a continuous mapping from a circle $S^1$ into the
order parameter space $G$, i.e., from the first homotopy group $\pi_1(G)$. Since
$\pi_1(SU(2)) = \pi_1(SO(3)) = Z_2$ and $\pi_1(U(1) \times S^2) = Z$, we
conclude that a spin-1/2 and a ferromagnetic spin-1 condensate can have
only vortices with a winding number equal to 1, whereas an antiferromagnetic 
spin-1
condensate can have vortices with winding numbers that are an arbitrary  
integer. Physically, this means that by traversing the boundary of the plane, 
the spinor can wind around the order parameter at most once or an arbitrary 
number of times, respectively.
This conclusion is identical to the one obtained prevously by
Ho \cite{jason1}.

If the vortex is nonsingular, however, the spinor $\zeta({\bf x})$ will be
identical everywhere on the boundary of the two-dimensional plane and it
effectively represents a mapping from the surface of a three-dimensional sphere
$S^2$ into the order parameter space. We then need to consider the second 
homotopy
group $\pi_2(G)$. For this we have that $\pi_2(SU(2)) = \pi_2(SO(3)) = 0$ and
$\pi_2(U(1) \times S^2) = Z$. Hence nonsingular or coreless vortices are only
possible for a spin-1 condensate with antiferromagnetic interactions. It
therefore appears that the nonsingular spin texture discussed in Ref.
\cite{jason1}, is topologically unstable and can be
continously deformed into the ground state by `local surgery' \cite{mermin}.   
        
We are now in a position to discuss point-like defects. Since the boundary of a 
three-dimensional gas is also the surface of a three-dimensional sphere, 
singular point-like defects are also determined by the second homotopy group 
$\pi_2(G)$. Such topological excitations thus only exists in the case of a 
spin-1 Bose gas with antiferromagnetic interactions. We call these excitations 
singular skyrmions, although in view of the work of 't Hooft and Polyakov it 
would be justified to call them monopoles \cite{mono}. For nonsingular 
point-like defects the spinor $\zeta({\bf x})$ will again be identical on the 
boundary of the three-dimensional gas. As a result, the configurations space is 
compactified to the surface of a four-dimensional sphere $S^3$ and we need to 
determine the third homotopy group $\pi_3(G)$. For this we find $\pi_3(SU(2)) = 
\pi_3(SO(3)) = \pi_3(U(1) \times S^2) = Z$. Hence nonsingular skyrmion 
excitations exists in all three cases.
 
{\it Skyrmion texture.} --- In view of the fact that magnetic gradients are 
needed to stabilize the spinor condensate in the case of a spin-1 Bose gas with 
antiferromagnetic interactions \cite{jason2}, we from now on consider only  
spin-1/2 and ferromagnetic spin-1 Bose gases, which have only nonsingular 
skyrmions as we have seen. Due to the absense of a core, the fluctuations in the 
density 
$\delta n({\bf x}) = n({\bf x}) - \langle n({\bf x}) \rangle$ will be small 
compared to the average density 
$\langle n({\bf x}) \rangle$ and the energy of the skyrmion 
can be determined by
\begin{eqnarray}
E[\zeta] = \frac{1}{2} \int d{\bf x} \int d{\bf x}'~ 
     \delta n({\bf x}) \chi^{-1}({\bf x},{\bf x}') 
                        \delta n({\bf x}')         \hspace*{0.3in} \nonumber \\
+ \int d{\bf x}~ n({\bf x})
     \left( \frac{\hbar^2}{2m} |\mbox{\boldmath $\nabla$} \zeta({\bf x})|^2 
         - \mu {\bf B} \cdot \langle {\bf S} \rangle({\bf x}) \right)~.      
\end{eqnarray}
Here, $\chi({\bf x},{\bf x}')$ is the static density-density response function 
defined by
\begin{eqnarray}
\frac{\hbar^2}{4m} 
  \left( - \mbox{\boldmath $\nabla$} \cdot 
  \left( \frac{1}{\langle n({\bf x}) \rangle} \mbox{\boldmath $\nabla$} \right)
    + 16\pi a \right) \chi({\bf x},{\bf x}')  \hspace*{0.5in} \nonumber \\
= \delta({\bf x}-{\bf x}')~,
\end{eqnarray}
$a$ is the appropriate scattering length, ${\bf B}$ is either a ficticious
magnetic field, caused by resonant rf-fields, or a real 
homogeneous magnetic field and $\mu$ is the corresponding magnetic moment of the 
atoms in the trap. Moreover, the spinor $\zeta({\bf x})$ can now be conveniently 
parametrized as
\begin{equation}
\zeta({\bf x}) 
   = \exp \left\{ - \frac{i}{S} \Omega^{\alpha}({\bf x}) S_{\alpha} \right\} 
\zeta^{\rm Z}~,
\end{equation}
where $S$ denotes the spin of the atoms and $\zeta^{\rm Z}$ is a constant spinor 
that minimizes the Zeeman energy, i.e., it obeys 
$\zeta^{\rm Z}_a = \delta_{aS}$ if we use 
the direction of the magnetic field as the quantization axis. 
   
This last equation explicitly shows that the topology of the order parameter 
space is a sphere of radius $\pi$ with opposite points on the surface 
identified. If we now assume that a maximally symmetric shape of the
skyrmion spin texture is allowed, we can take
$\mbox{\boldmath $\Omega$}({\bf x}) = {\bf x} \omega(x)/x 
                                              \equiv \hat{\bf x} \omega(x)$,
with $\omega(x)$ a monotonically decreasing function obeying $\omega(0) = 2\pi$ 
and 
$\lim_{x \rightarrow \infty} \omega(x) = 0$.
For this ansatz we see that by traversing the whole 
configuration space, we exactly cover the order parameter space once. The above 
spin texture therefore corresponds to a skyrmion with a topological winding 
number
\begin{equation}
W = \frac{3}{4\pi^4} \int d{\bf \Omega} 
  = \frac{1}{8\pi^4} \int d{\bf x} \epsilon^{ijk} \epsilon_{\alpha\beta\gamma}
      \partial_i \Omega^{\alpha} \partial_j \Omega^{\beta}
      \partial_k \Omega^{\gamma}
\end{equation}
equal to 1 and located at the center of the trap.

To obtain the precise spin texture of the skyrmion we have to find the function 
$\omega(x)$ that minimizes the energy. In general this leads, due to the  
nonlocality of the density-density response function, to a complicated nonlinear 
integro-differential equation, which can only be solved numerically. However, we 
can gain a lot of physical insight in the size and stability of the skyrmion by 
the following analysis. Considering first the ideal case of zero magnetic field 
${\bf B}$ and solving for the density fluctuations induced by the spin texture, 
we have to minimize the gradient energy
\begin{eqnarray}
E^{\rm grad}[\zeta] = \int d{\bf x}~ \langle n({\bf x}) \rangle
      \frac{\hbar^2}{2m} |\mbox{\boldmath $\nabla$} \zeta({\bf x})|^2
             \hspace*{0.8in}  \nonumber \\
- \frac{\hbar^4}{8m^2} \int d{\bf x} \int d{\bf x}'~ 
      |\mbox{\boldmath $\nabla$} \zeta({\bf x})|^2 \chi({\bf x},{\bf x}') 
      |\mbox{\boldmath $\nabla$} \zeta({\bf x}')|^2~.  
\end{eqnarray}
For a skyrmion of size $\lambda$ in the center of the trap this energy can be 
estimated to be given by
\begin{eqnarray}
E^{\rm grad}(\lambda) \simeq \langle n({\bf 0}) \rangle \frac{2\pi\hbar^2}{m}
   \left( \lambda - \frac{3 \lambda }{(\lambda/\xi)^2 +1} \right)~,
\end{eqnarray}
with $\xi = (16\pi |a| \langle n({\bf 0}) \rangle)^{-1/2}$ the typical coherence 
length of the condensate. It has a minimum when $\lambda/\xi$ is equal to 
$((\sqrt{33}-5)/2)^{1/2} \simeq 0.61$. Therefore, our maximally symmetric
ansatz is indeed justified, even for an anisotropic trap, as long as the 
coherence
length is much smaller than the size of the condensate, i.e., we are 
in the
so-called Thomas-Fermi limit.     

For this result we have assumed the scattering length $a$ to be positive. In 
that case the skyrmion is thus stabilized by the fact that gradients in the spin 
texture on the one hand lead to an increase in the average kinetic energy of the 
spinor condensate, but one the other hand also lead to a reduction of the 
density, which for repulsive interactions results in a decrease in the average 
interaction energy. For a negative scattering length, gradients in the spin 
texture lead to an enhancement of the density, but this, due to the attractive 
interactions, also reduces the average interaction energy. Therefore, we expect 
skyrmions also to be stablilized in that case. Indeed, using the same estimates 
as before, we find that $E^{\rm grad}(\lambda)$ is now minimized for 
$\lambda/\xi$ equal to $((\sqrt{33}+5)/2)^{1/2} \simeq 2.3$. However, due to the 
intrinsic instability of a condensate with attractive interactions \cite{henk1}, 
this implies that the size of the skyrmion is essentially equal to the size of 
the condensate itself and that its spin texture will be strongly affected by 
finite-size effects. In particular, it in general becomes anisotropic. 

Note finally, that the above mechanism for the stability of the skyrmion 
excitations is quite different from that in the quantum Hall effect near filling 
fraction $\nu = 1$. There 
skyrmions are electrically charged and their size is determined by a competition 
between the Coulomb and Zeeman interactions \cite{Sondhi}. In the case of a 
spinor Bose-Einstein condensate, the Zeeman interaction plays a much less 
important role and for small magnetic fields $B \ll 8\pi |a|\hbar^2\langle 
n({\bf 0}) \rangle/m\mu$ only directs the spins at large distances of the 
skyrmion. Larger magnetic fields will tend to decrease the size of the skyrmion 
so as to reduce the cost in Zeeman energy.

{\it Skyrmion dynamics.} --- The most important dynamical variable of the 
skyrmion arises from the fact that the Euler-Lagrange equations for the skyrmion 
spin texture is invariant under a space independent rotation of the average 
local spin $\langle {\bf S} \rangle({\bf x})$ around the magnetic field 
direction $\hat{\bf B}$ \cite{rene}. Mathematically, this means that if 
$\zeta^{\rm sk}({\bf x})$ describes a skyrmion, then 
$\exp\{-i\vartheta \hat{\bf B} \cdot {\bf S}\} \zeta^{\rm sk}({\bf x})$ 
describes also a skyrmion with the same winding number and energy.
The dynamics of the variable 
$\vartheta(t)$ associated with this symmetry is determined by the full action 
for the spin texture $S[\zeta] = \int dt (T[\zeta] - E[\zeta])$, which contains 
the time-derivative term
\begin{equation}
T[\zeta] = \int d{\bf x}~
  n({\bf x},t) 
     \zeta^*({\bf x},t) i\hbar \frac{\partial}{\partial t} \zeta({\bf x},t)~.
\end{equation}

Hence, subsituting
$\zeta({\bf x},t)
    = \exp\{-i\vartheta(t) \hat{\bf B} \cdot {\bf S}\} \zeta^{\rm sk}({\bf x})$
and making use of the conservation of total particle number $N$ to introduce
the change of the average local spin projection on the magnetic field
$\langle \Delta S_{||} \rangle({\bf x})
    = \hat{\bf B} \cdot \langle {\bf S} \rangle({\bf x}) - S$
induced by the skyrmion,
we obtain apart from an unimportant boundary term that
\begin{equation}
T[\zeta] = \hbar  \frac{\partial \vartheta(t)}{\partial t} \int d{\bf x}~
  n({\bf x},t) \langle \Delta S_{||} \rangle({\bf x})~.
\end{equation}
Taking now again the density fluctuations into account, we finally find that
the dynamics of the rotation angle $\vartheta(t)$ is determined by the action
\begin{equation}
\label{action}
S[\vartheta] = \int dt~
  \left\{ \frac{\partial \vartheta(t)}{\partial t}
             \hbar \langle \Delta S^{\rm tot}_{||} \rangle
    + \frac{1}{2} I \left( \frac{\partial \vartheta(t)}{\partial t} \right)^2
  \right\}~,
\end{equation}
where $\langle \Delta S^{\rm tot}_{||} \rangle$ is the change of the total
spin along the magnetic field direction and the `moment of inertia' of the
skyrmion equals
\begin{equation}
I = \hbar^2 \int d{\bf x} \int d{\bf x}'~
       \langle \Delta S_{||} \rangle({\bf x})
          \chi({\bf x},{\bf x}') \langle \Delta S_{||} \rangle({\bf x}')~.  
\end{equation}

The importance of this result is twofold. First, from the action
in Eq.~(\ref{action}) we see that at the quantum
level the hamiltonian for the dynamics of the wave function
$\Psi(\vartheta,t)$ becomes
\begin{equation}
H = \frac{1}{2I} \left( \frac{\hbar}{i}
                          \frac{\partial}{\partial \vartheta}
                        - \hbar \langle \Delta S^{\rm tot}_{||} \rangle
                 \right)^2~.
\end{equation}
Therefore, the ground state wave function is given by
$\Psi_0(\vartheta) = e^{iK\vartheta}/\sqrt{2\pi}$, with $K$ an integer that
is as close as possible to $\langle \Delta S^{\rm tot}_{||} \rangle$.
In this way we thus recover the fact that according to quantum mechanics the
total number of spin-flips associated with the skyrmion texture must be an
integer. More precisely, we have actually shown that the many-body wave
function describing the skyrmion is an eigenstate of the operator
$\hat{\bf B} \cdot {\bf S}^{\rm tot}$ with eigenvalue $NS - K$. Note that
physically this is equivalent to the way in which `diffusion' of the overall
phase of a Bose-Einstein condensate leads to the conseration of particle
number \cite{li,henk2}. 

Furthermore, the existence of this internal degree of freedom
becomes especially important when we deal with more than one skyrmion in
the condensate. In that case every skyrmion can have its own orientation
and we expect the interaction between two skyrmions to have a
Josephson-like contribution proportional to $\cos(\vartheta_1-\vartheta_2)$.
As a result the phase diagram of a gas of skyrmions can become extremely
rich and contain both quantum as well as classical, i.e., nonzero
temperature, phase transitions \cite{rene}. In this context, it is interesting
to mention two important differences with the situation in the quantum Hall 
effect. First, the fact that the spin projection $K$ of the skyrmion is an 
integer shows that these excitations have an integer spin and are therefore
bosons \cite{FR}. In the quantum Hall case the skyrmions are fermions with 
half-integer spin, due to the presence of a topological term in the action 
$S[\zeta]$ for the spin texture \cite{wilczek}.
Second, in a spinor Bose-Einstein condensate the skyrmions are not pinned by 
disorder and are in principle free to move. Both differences 
will clearly have important consequences for the many-body physics of a 
skyrmion gas. 

Focussing again on the single skyrmion dynamics, we can make the last remark 
more quantitative by using the ansatz 
$\zeta({\bf x},t)= \zeta^{\rm sk}({\bf x}-{\bf u}(t))$ for the texture of a 
moving
skyrmion, which is expected to be accurate for small velocities 
$\partial {\bf u}(t)/\partial t$. Considering for illustrative purposes also 
only the case of a maximally symmetric skyrmion near the center of the trap
with $u/\xi \ll 1$, we find in a similar way as before that the action for
the center-of-mass motion of the skyrmion becomes
\begin{equation}
S[{\bf u}] = \int dt~ 
 \frac{1}{2} M \left( \frac{\partial {\bf u}(t)}{\partial t} \right)^2~,   
\end{equation}
where the mass, which in the more general anisotropic case is of course a 
tensor, is 
now
simply given by 
\begin{equation}
M = \frac{m^2}{3} \int d{\bf x} \int d{\bf x}'~ \chi({\bf x},{\bf x}')
       \langle {\bf v}_{\rm s} \rangle({\bf x}) \cdot
           \langle {\bf v}_{\rm s} \rangle({\bf x}')  
\end{equation}
in terms of the superfluid velocity of the spinor condensate 
$\langle {\bf v}_{\rm s} \rangle({\bf x}) 
   \equiv -i\hbar {\zeta^{\rm sk}}^*({\bf x}) \mbox{\boldmath $\nabla$} 
                                                   \zeta^{\rm sk}({\bf x}) /m$.
The skyrmions thus indeed behave in this respect as ordinary particles. It
should be noted that in contrast to Eq.~(\ref{action}) there appears no term
linear in $\partial {\bf u}(t)/\partial t$ in the action $S[{\bf u}]$. This is
a result of the fact that we have performed all our calculations at zero 
temperature.
In the presence of a normal component, we anticipate the appearance of such a
linear term with an imaginary coefficient. This will lead to damping of the
center-of-mass motion of the skyrmion.

I would like to thank Michiel Bijlsma, Gerard 't Hooft, David Olive, and Jan 
Smit for useful discussions. This work is supported by the Stichting 
Fundamenteel Onderzoek der Materie (FOM), which is financially supported by the 
Nederlandse Organisatie voor Wetenschappelijk Onderzoek (NWO).

\end{multicols}

\begin{references}
\bibitem{scat} In principle, this is only true if the inter and intra
               scattering lengths of the two condensate components are equal,
               which is a good approximation for $^{87}$Rb \cite{JILA2}.
\end{references}
\end{document}